\documentclass[twocolumn,prd,superscriptaddress,showpacs,amsmath,amssymb]{revtex4-1}

\usepackage{graphicx}
\usepackage{dcolumn}
\usepackage{bm}
\usepackage{multirow}
\usepackage{subfigure}
\usepackage{epsfig}

\begin{document}

\title{Maximum Likelihood Signal Extraction Method Applied to 3.4 years of CoGeNT Data}

\def\PNNL{Pacific Northwest Laboratory, Richland, WA 99352, USA}
\def\UC{Kavli Institute for Cosmological Physics and Enrico Fermi Institute, University of Chicago, Chicago, IL 60637, USA}
\def\CANBERRA{CANBERRA Industries, Meriden, CT 06450, USA}
\def\UW{Center for Experimental Nuclear Physics and Astrophysics and Department of Physics, University of Washington, Seattle, WA 98195, USA}

\author{C.E.~Aalseth} \affiliation{\PNNL}
\author{P.S.~Barbeau} \altaffiliation{Present address: Department of Physics, Duke University, Durham, NC 27708, USA} \affiliation{\UC} 
\author{J.~Diaz Leon} \affiliation{\UW}
\author{J.E.~Fast} \affiliation{\PNNL}
\author{T.W.~Hossbach} \affiliation{\PNNL}
\author{A.~Knecht} \altaffiliation{Present address: Paul Scherrer Institut (PSI), Villigen PSI, Switzerland}\affiliation{\UW}
\author{M.S.~Kos} \email{Electronic address: marek.kos@pnnl.gov}\affiliation{\PNNL}
\author{M.G.~Marino} \altaffiliation{Present address: Physics Department, Technische Universit\"at M\"unchen, Munich, Germany}\affiliation{\UW}
\author{H.S.~Miley} \affiliation{\PNNL}
\author{M.L.~Miller} \altaffiliation{Present address: Cloudant - West Coast, 209 1/2 1$^{\textrm{st}}$ Ave S,  Seattle, WA 98104}\affiliation{\UW}
\author{J.L.~Orrell}  \affiliation{\PNNL}


\date{\today}

\begin{abstract}
CoGeNT has taken data for over 3 years,  with 1136 live days of data accumulated as of April 23, 2013. We report on the results of a maximum likelihood analysis to extract any possible dark matter signal present in the collected data. The maximum likelihood signal extraction uses 2-dimensional probability density functions (PDFs) to characterize the anticipated variations in dark matter interaction rates for given observable nuclear recoil energies during differing periods of the Earth's annual orbit around the Sun. Cosmogenic and primordial radioactivity backgrounds are characterized by their energy signatures and in some cases decay half-lives. A third parameterizing variable -- pulse rise-time -- is added to the likelihood analysis to characterize slow rising pulses described in prior analyses. The contribution to each event category is analyzed for various dark matter signal hypotheses including a dark matter standard halo model and a case with free oscillation parameters (i.e., amplitude, period, and phase). The best-fit dark matter signal is in close proximity to previously reported results.  We find that the significance of the extracted dark matter signal remains well below evidentiary at 1.7 $\sigma$. 
\end{abstract}



\maketitle

\section{Introduction}

The CoGeNT detector has operated stably for over three years at the Soudan Underground Laboratory. Prior publications~\cite{CogPRLa,CogPRLb,CogPRD} have analyzed data from the CoGeNT detector testing the collected data for any signature of dark matter interactions. Those prior results have shown a preference for an excess of events above the expected background. If this excess is treated as a dark matter signal, a best fit is found for a low mass ($\sim$10~GeV/c$^{2}$) dark matter particle with large ($\sim$10$^{42}$~cm$^{2}$) interaction cross-section. These results are based on analyzing the CoGeNT data for both the recoil nuclear energy spectral signature~\cite{CogPRLa} and separately the expected temporal variation of the event rate with an annual period and summer to winter phase~\cite{CogPRLb}. In all cases the statistical significance of these observations are just beneath evidentiary (i.e., $3\sigma$). While the significance of the CoGeNT results (and other recent findings~\cite{CDMSSi}) do not rise to the level of discovery, the possible detection of the first non-Standard Model particle should not be blithely ignored nor accepted without irrefutable evidence. Thus a detailed analysis of the released 3.4~year CoGeNT data set is pursued. The objective is to use models of the backgrounds present in the CoGeNT data set -- parameterized by energy and time -- to test for a possible dark matter signal using a fully unbinned maximum likelihood signal extraction method. The principle underlying this method is generic and applicable in current and future dark matter searches, thus a focus on reporting the methodology is presented in this article. Finally, this maximum likelihood methodology provides a quantifiable way to test for the effect of systematic uncertainties associated with understanding of the signal and background distributions.

\begin{figure}[!htbp]
    \begin{center}
       \subfigure[~0.5-0.7~keV]{\epsfig{file=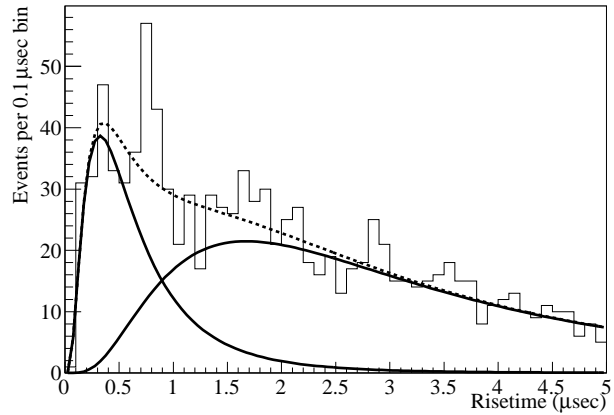, width=9cm}}
       \subfigure[~2-3~keV]{\epsfig{file=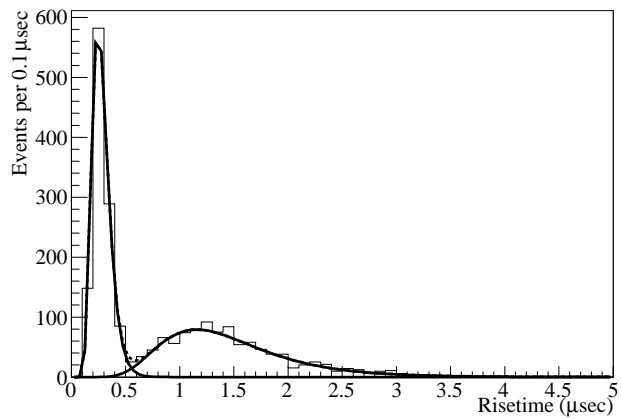, width=9cm}}       
       \subfigure[~4-5~keV]{\epsfig{file=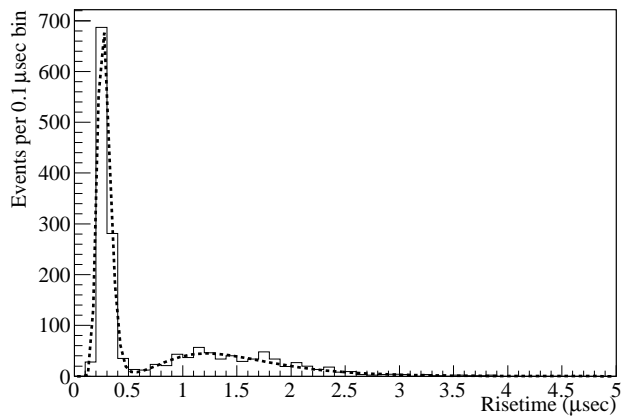, width=9cm}}       
       \caption{The rise-time distributions for various energy ranges from the 3.4~year data set. The distributions are fit with pairs of log-normal distributions for the fast and slow rise-time distributions.  The solid curves are the individual fitted log-normal functions and the dashed curves are the combined functions.  For the panel (c),  the solid and dashed curves overlap completely therefore we only show the combined function.}
       \label{fastslowfits}
    \end{center}
\end{figure}

Of particular interest in this analysis is addressing a background having a similar energy spectral shape to that expected from dark matter induced nuclear recoil interactions (e.g., a falling exponential as a function of increasing energy). This background is due to energy depositions in the high voltage contact surface layers of the germanium crystal.  The surface event pulses on average have longer rise-time than bulk events~\cite{Sakai,ryan,CogPRD}.  In recent published analysis, the fast and slow pulse rise-time distributions were shown to be well approximated by log-normal functions, a model qualitatively justified by the impact of electronic signal noise on the determination of individual pulse rise-times~\cite{CogPRD}. The separation between fast and slow pulses is very good at higher energies, with the separation becoming worse at lower energies as the risetime distributions broaden.  Figure~\ref{fastslowfits} shows distributions of the log-normal functions fitted to the data in three energy ranges: 0.5--0.7~keV, 2--3~keV, and 4--5~keV (additional energy ranges are shown in Ref.~\cite{CogPRD}).  It is clear the separation becomes very poor at low energies and is the primary source of uncertainty in this analysis.  Panels b and c of figure~\ref{fastslowfits} show the log-normal functions are good approximations for the fast and slow pulse distributions at higher event energies.

In the analysis presented here an unbinned maximum likelihood signal extraction is performed on 1136 live-days of data seeking to extract any possible dark matter signal in the presence of backgrounds. The backgrounds included are the L-shell x-ray peaks (constrained by K-shell x-ray peak intensities), a muon-induced neutron background (constrained by analysis of the cosmic ray veto panel data), a flat Compton distribution from external gamma rays (constrained by the flat continuum under the K-shell x-ray peaks), and a surface event background from external gamma rays (studied as a dominant contributor to the systematic uncertainty).  There are no rise-time cuts applied in this analysis, instead probability density functions (PDFs) in rise-time for bulk and surface events are used to determine the relative contributions from these categories of events. As the contribution to each event population is dependent on the details of the model PDF, effort is applied to understanding the effect of systematic uncertainties associated with the PDF models used.

\section{Likelihood Signal Extraction Method}

In a maximum likelihood signal extraction, the most likely level of contribution to the data set from each of the known backgrounds and any possible signal are estimated (See Ref.~\cite{PDG} Ch. 35 and 36). For likelihood signal extraction applied to the CoGeNT data set, the total probability for any event is given by
\begin{equation}
\mathcal{P}_{i} = \alpha_{\textrm{L}}P_{\textrm{L}} + \alpha_{\textrm{n}}P_{\textrm{n}} + \alpha_{\textrm{flat}}P_{\textrm{flat}} + \alpha_{\textrm{surf}}P_{\textrm{surf}} + \alpha_{\chi}P_{\chi} \label{pdf1},
\end{equation}
\noindent where $P_{\textrm{L}}$ represents the probability that the event is one of the L-shell x-rays, $P_{\textrm{n}}$ is the probability that the event being is a neutron, $P_{\textrm{flat}}$ is the probability the the event is a part of a ``flat'' linear background, $P_{\textrm{surf}}$ is the probability that the event is a surface event, and finally $P_{\chi}$ is the probability that the event is due to a dark matter interaction.  The various $\alpha$'s are the size of the contributions from each of the signal groups in the data.  There are ten separate background signals in the maximum likelihood extraction.  For the cosmogenic backgrounds we have the L-shell x-rays for $^{68}$Ge, $^{68}$Ga, $^{65}$Zn, $^{73,74}$As, $^{56,57,58}$Co, $^{55}$Fe, and $^{54}$Mn.  The other backgrounds are muon-induced neutrons,  the bulk Compton continuum, and surface gamma rays.  

The ten background PDFs are parameterized as
\begin{equation}
P_{\textrm{bkg}}(E,rt,t) = P(rt,E) \times P(t). 
\label{pdfparam0}
\end{equation}
Using the product rule,  equation~\ref{pdfparam0} can be written using the conditional probability distribution function, $P(rt | E)$:
\begin{equation}
P_{\textrm{bkg}}(E,rt,t) = P(rt | E) \times P(E) \times P(t), 
\label{pdfparam}
\end{equation}
\noindent which is the form we use in our signal extraction.  The probability distribution $P(E)$ describes the detector's expected spectral energy response to the background, $P(t)$ describes the expected temporal variation of the background with time, and $P(rt|E)$ is a conditional probability distribution function describing the detector's expected pulse rise-time response for a given event energy. For the case of the standard halo model,  the dark matter signal PDF is parameterized as 
\begin{equation}
P_{\chi}(E,t,rt) =  P(rt | E) \times P(E,t).
\label{pdfparam2}
\end{equation}
The probability distribution $P(E,t)$ describes the detector's energy-spectral and temporal response to dark matter interactions as a fully two-dimensional PDF. The probability distribution $P(rt|E)$ has the same meaning as described above for backgrounds. For the case of a signal extraction with fully free WIMP oscillation parameters, equation~\ref{pdfparam2} is parameterized exactly like equation~\ref{pdfparam}.   In these PDFs the variables $E$, $t$, and $rt$ are energy, time (since 3 December 2009), and pulse rise-time, respectively.  To extract the signal contribution we minimize the log of the likelihood function
\begin{equation}
\mathcal{L}_{\log} = -2 \sum_{i=0}^{N} \log(\mathcal{P}_{i}) \label{eqn:like}.
\end{equation}
where the summation is over all $N$ events selected (see Sec.~\ref{sec:SigEx} for event selection discussion).

\subsection{Constraints in the Likelihood fit}
It is possible additional information is available regarding the relationships between the different PDFs that is not or need not be represented in the three-variable parameter space of $E$, $t$, and $rt$ for which the event populations have constructed PDFs. This information can include expected intensity ratios between the event populations or secondary measurements that limit the range of contribution from a particular event category. These 'external' constraints can be included in the maximum likelihood signal extraction. Equation \ref{econstraint} describes how this is done if $\alpha$ is the fitted value, $x$ is the external measurement of that value, and $\sigma_{x}$ is the uncertainty on the external measurement.
\begin{eqnarray}
p_{c} & = &\frac{1}{\sqrt{2 \pi} \sigma_{x}}e^{\frac{1}{2}\left(\frac{\alpha-x}{\sigma_{x}}\right)^{2}} , \\
\mathcal{L}^{c}_{\log} & = & \mathcal{L}_{\log} - 2\log(p_{c}).
\label{econstraint}
\end{eqnarray}
\noindent The constrained log-likelihood function, $\mathcal{L}^{c}_{\log}$, is the function we minimize in our signal extraction.

\section{Probability Density Function Descriptions}

Each of the PDFs for the background contributions to the CoGeNT data set are described in this section. In general, one should consider the maximum likelihood signal extraction as performed on data from the high-gain energy channel (CH1)~\cite{CogPRD}, compromised of events falling in the 0.5-3.0~keV energy range. However, where it is possible to provide constraints, other channels of information from the CoGeNT data acquisition~\cite{CogPRD} are used. These details are described in this section. The dark matter signal PDF, $P_{\chi}$, is described alongside the signal extraction results in a later section as several forms of $P_{\chi}$ are explored in the analysis presented in this article.

\subsection{Pulse rise-time distributions, $P(rt | E)$}
\begin{figure}[!htbp]
    \begin{center}
       \subfigure[~Bulk events]{\epsfig{file=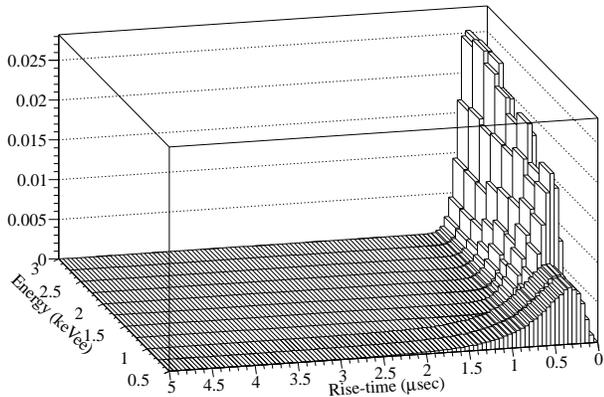,width=0.5\textwidth}}
       \subfigure[~Surface events]{\epsfig{file=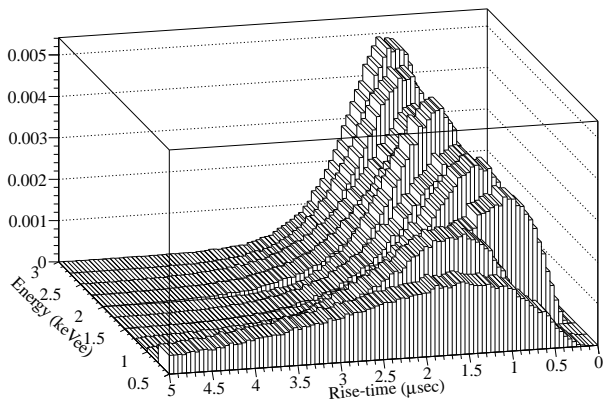,width=0.5\textwidth}}
       \caption[Two-dimensional conditional PDFs used in the signal extraction]{Two-dimensional conditional PDFs for event pulse rise-time in 0.25~keV-wide energy bins. Derived from the collected data, these PDFs are used to describe the bulk and surface event populations in the signal extraction analysis.}
       \label{fig:bulksurfpdfs}
    \end{center}
\end{figure}

The most effective way to separate surface from bulk events is through pulse rise-time~\cite{CogPRD}. The bulk events have,  on average,  faster pulse rise-times than surface events.  In the maximum likelihood signal extraction we use the two dimensional PDFs shown in figure~\ref{fig:bulksurfpdfs} to determine the correlation between pulse rise-time and event energy.  These PDFs are expressed as $P(rt | E)$ in equations~\ref{pdfparam0} and~\ref{pdfparam2}. The rise-time PDFs are binned in 0.25 keV wide energy bins and 0.1~$\mu$sec wide rise-time bins.  The rise-time distributions for each energy bin are all normalized to unity.  Therefore these PDFs provide for the variation of the rise-time distribution as a function of energy. The conditional probability distribution functions shown in figure~\ref{fig:bulksurfpdfs} are generated by fitting log-normal functions to the rise-time distributions of the collected 1136~live-day data set when placed into 0.25~keV-wide energy bins.  Ideally we would like to use calibration data to determine the shape of these conditional PDFs. However, until these calibrations are collected in future detector characterization runs, we use the existing PDFs created from the collected data to demonstrate the viability of the maximum likelihood signal extraction method.

\subsection{L-shell X-rays, $P_{\textrm{L}}$}
\begin{figure}[!htp]
   \includegraphics[width=0.5\textwidth]{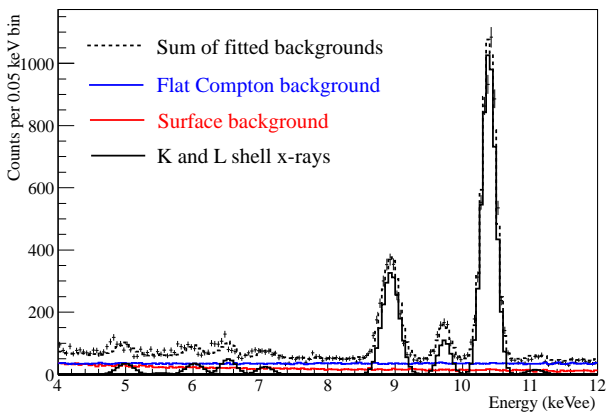}
   \caption{\label{fig:HE} The fitted energy distributions from the signal extraction performed in the 4 - 12 keVee region compared to the data.}
 \end{figure}

Cosmogenic activation of the germanium crystal takes place when the crystal is on the surface during fabrication. This results in radioactive isotopes present uniformly through-out the germanium crystal (isotopes labeled in figure~3 of Ref.~\cite{CogPRLa}). The presence of electron capture decay isotopes are identified by the characteristic K-shell x-rays emitted and immediately reabsorbed in the germanium crystal. The K-shell x-rays in the high energy data channel are fit directly to determine their intensity and as a check for the appropriate isotope's decay half-life.  Figure~\ref{fig:HE} shows the result of the signal extraction in the high-energy region compared to the data in the high energy channel.  The fit K-shell contributions in the high-energy region are used as constraints on the L-shell peaks for the signal extraction on the low-energy data.

\begin{figure}[!htbp]
   \includegraphics[width=0.5\textwidth]{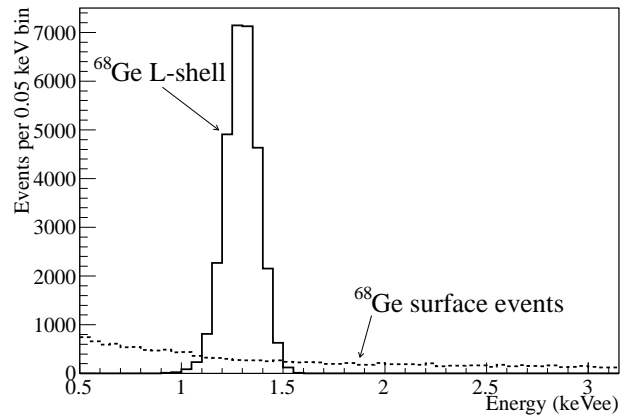}
   \caption{\label{gebulktrans} Energy distributions taken from the Monte Carlo simulation of $^{68}$Ge L-shell x-rays in the bulk and for both K-shell and L-shell x-rays in the surface transition region.  The surface events above threshold are predominantly due to K-shell x-rays.}
 \end{figure}

The L-shell energy PDFs, $P(E)$, are Gaussian peaks with resolutions taken from the energy resolution versus energy equation given in~\cite{CogPRLerrata}. The mean energies for each L-shell x-ray peak are taken from~\cite{TORI}, and the K- to L-shell intensity ratios that are used are from~\cite{Bahcall}. The time evolution of the individual L-shell x-ray contribution PDFs, $P(t)$, is taken from the standard isotopic half-life values~\cite{TORI}. For decays occurring in the small ($<$14\% by volume) region of partial charge collection near the detector surface, the reduced charge collection results in most of these events being degraded to below the analysis energy threshold. However, some of the K-shell x-rays in the surface transition region will be above threshold. See section~\ref{subsec:surface} for a description of how degraded energy values are calculated for events in the partial charge collection surface region.  Figure~\ref{gebulktrans} shows the relative scale of the L-shell peak events relative to the K-shell surface events for the case of $^{68}$Ge x-rays. In the forthcoming signal extraction we include PDFs for both the bulk L-shell peaks and the associated surface event contribution, which is mostly from the K-shell x-rays.
\noindent However, we do not fit for the size of the K-L shell surface event contributions,  but rather scale these PDFs to be the size of the corresponding L-shell contribution $\times$ the L-shell to surface component ratio that is given by the Monte Carlo (figure~\ref{gebulktrans} shows this ratio for $^{68}$Ge). 

\subsection{Neutrons, $P_{\textrm{n}}$}

The neutron PDF represents neutrons produced by cosmic ray muons passing through the CoGeNT shield. Prior analysis~\cite{CogPRD} presented the results from a GEANT Monte Carlo simulation of this process and the resulting nuclear recoil event energy spectrum in germanium. That simulation provides the shape of the $P(E)$ PDF for these neutrons. The temporal modulation amplitude for muons has been measured to be $\sim$2\% at the Soudan Underground Laboratory~\cite{MINOSdata}. For the purpose of this analysis the time PDF, $P(t)$, is treated as \emph{constant} in time. This is sensible when the modulation of the muon rate is a $\sim$2\% effect on a small contribution to the total CoGeNT data set; this is estimated at a O(0.1\%) effect in the CoGeNT data~\cite{CogPRD}.

\begin{figure}[!htbp]
   \includegraphics[width=0.5\textwidth]{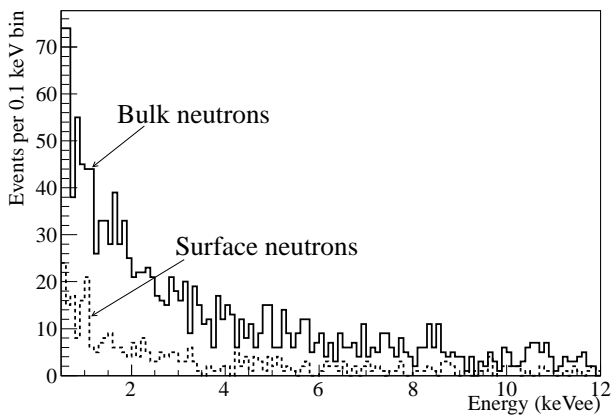}
   \caption{\label{neutronsbs} Monte Carlo simulation of the muon-induced neutron energy distributions in the bulk and surface transition regions.}
 \end{figure}

The final PDF needed to describe the neutron event population is the risetime-energy PDF, $P(rt|E)$. Neutrons will interact uniformly throughout the germanium crystal. A small fraction will interact in the previously mentioned surface transition region. Further, application of a partial charge collection model (See Section~\ref{subsec:surface}) in this region means even fewer of the low energy nuclear recoils produced by this class of muon induced neutrons will produce events that are above the energy threshold of the analysis. To take into account the neutrons depositing their energy in the transition region,  there are two categories of neutron PDFs, bulk neutrons and transition region neutrons.  The relative scale between these is fixed in the maximum likelihood signal extraction and this scale is determined from Monte Carlo.  Figure~\ref{neutronsbs} shows the neutron energy distribution for both surface and bulk energy depositions and their relative scale.  The muon-induced neutron Monte Carlo is computationally intensive,  since it starts with muons impinging on the CoGeNT shield from the cavern and propagates all secondary particles produced.  This limits the statistics in the PDFs shown in figure~\ref{neutronsbs}.  We have used both these PDFs and functional form approximations to these PDFs in the signal extraction.  Both PDF types produce very similar results. 

A constraint on the muon-induced neutron background contribution is inferred from measurements of the germanium detector coincidences with the muon-veto panel rate and a simulation of muon-induced events in the CoGeNT shielding. A prior analysis~\cite{CogPRD} examined the coincidence rate between the muon-veto panel array and the germanium detector. The prediction for this coincidence rate is based on the known muon rate at the Soudan Underground Laboratory~\cite{MINOSdata} and was shown to closely match the actual measured coincidence rate (See Fig.~17 in Sec. IV.A. of Ref.~\cite{CogPRD}). The measured germanium-veto coincidences rate per day of 0.67$\pm$0.12 is used as a constraint on the extracted number of neutron events.  The constraint is applied via equation~\ref{econstraint} in the maximum likelihood minimization.  This constraint is important because the nuclear recoil energy spectrum shape from muon-induced neutron scatters can potentially closely mimic the energy recoil spectrum shape from dark matter interactions. Together this rate constraint and the time components of the PDFs, $P(t)$, should in principle separate the muon-induced neutron contribution from any dark matter interaction contribution. Later in this analysis the impact of freeing this constraint on the muon-induced neutron contribution is studied.

\subsection{Flat Background, $P_{\textrm{flat}}$}\label{subsec:flat}
The ``flat'' linear continuum background PDF represents high energy gamma-ray Compton-scattering interactions in the bulk of the crystal, mostly due to uranium and thorium contamination in the surrounding materials.  The energy PDF, $P(E)$, for this background is assumed to be constant in the 0.5-12~keVee region. This has been verified by Monte Carlo simulations of uranium and thorium chain gamma rays~\cite{CogPRD}.  Figure~\ref{surfbulktrans} shows the energy distribution,  determined from simulations, of the uranium and thorium chain gammas depositing their energy in the bulk of the crystal.  For the time component of this background, $P(t)$, three different PDF types were tested: constant in time,  modulating in time,  and decaying over time. As the flat continuum events represent events taking place in the bulk of the crystal the $P(rt|E)$ PDF for these events is represented only by fast rising bulk events (i.e., Fig.~\ref{fig:bulksurfpdfs}~(a)).

\subsection{Surface Events, $P_{\textrm{surf}}$}\label{subsec:surface}

\begin{figure}[!htbp]
   \includegraphics[width=0.5\textwidth]{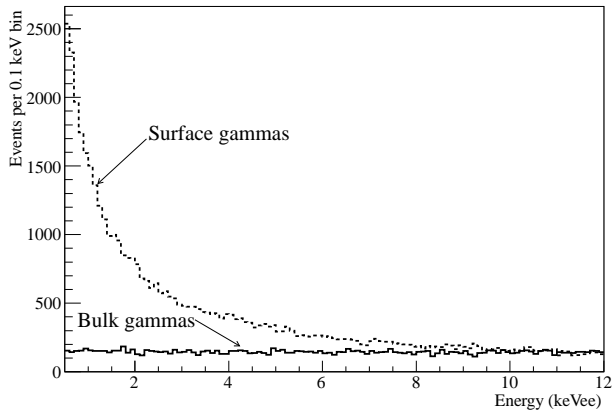}
   \caption{\label{surfbulktrans} The energy distributions of the surface and bulk events taken from the Monte Carlo simulation of the external uranium and thorium chain gammas.}
 \end{figure}

The same Monte Carlo employed above to investigate the energy spectrum due to uranium and thorium gamma-rays Compton-scattering in the bulk of the germanium crystal (See Section~\ref{subsec:flat}) is used to study the impact of the partial charge collection transition region of the germanium detector. In this partial charge collection region analysis we include the dead and transition regions on the high voltage contact surface of the germanium crystal.  The assumptions for the thickness of these regions were taken from~\cite{CogPRLa,ryan,paddy}. We assume that the thickness of both the dead and transition regions is 1 mm,  since analysis on the other detectors has yielded similar dimensions.   From the Monte Carlo simulation we get the energy distribution for the events depositing their energy in the surface transition regions.  In the Monte Carlo we use the 2-parameter sigmoid functional form for the charge collection efficiency as a function of depth into the Ge crystal given in~\cite{CogPRD}.  In this way we are able to generate the energy PDF, $P(E)$, for the surface events.  The charge collection efficiency decreases very rapidly with distance into the transition region from the bulk,  therefore large energy depositions in the surface transition region can have resulting energies that are within our signal region.  Figure~\ref{surfbulktrans} shows the results of this simulation for both surface-only and bulk-only events. 
\noindent The relative scale between the surface and bulk events in figure~\ref{surfbulktrans} comes directly from the Monte Carlo and is determined by the thickness of the transition region and the charge collection efficiency profile. 

The Monte Carlo energy distribution for surface events shown in figure~\ref{surfbulktrans} is compared to a population of surface event data in the next section.  Figure~\ref{mccompsurf} shows a comparison of the Monte Carlo simulation of surface events to the data.  The Monte Carlo does not give us the pulse rise-time distribution of the surface events.  This surface event population is the single most difficult background to address in the analysis so additional attention is given to the PDF modeling of this event class in the following section.

As with the bulk gamma-ray events described in Section~\ref{subsec:flat}, the temporal variation $P(t)$ of the event rate for the surface events is described in correspondence with the three different PDF types of constant in time,  modulating in time,  and decaying over time.

The slow-pulse surface event distribution and the Compton continuum extend into the high-energy region, therefore fitting in the high-energy region can also place constraints on those contributions. This is done via the constrained log-likelihood function,    $\mathcal{L}^{c}_{\log}$,  given in equation~\ref{econstraint}. 


\section{Systematic Studies of the Surface Event Distribution}\label{sec:SysStudy}
\begin{figure*}[!htbp]
    \begin{center}
       \subfigure[~Rise-time selection. All events to the right of each line.]{\epsfig{file=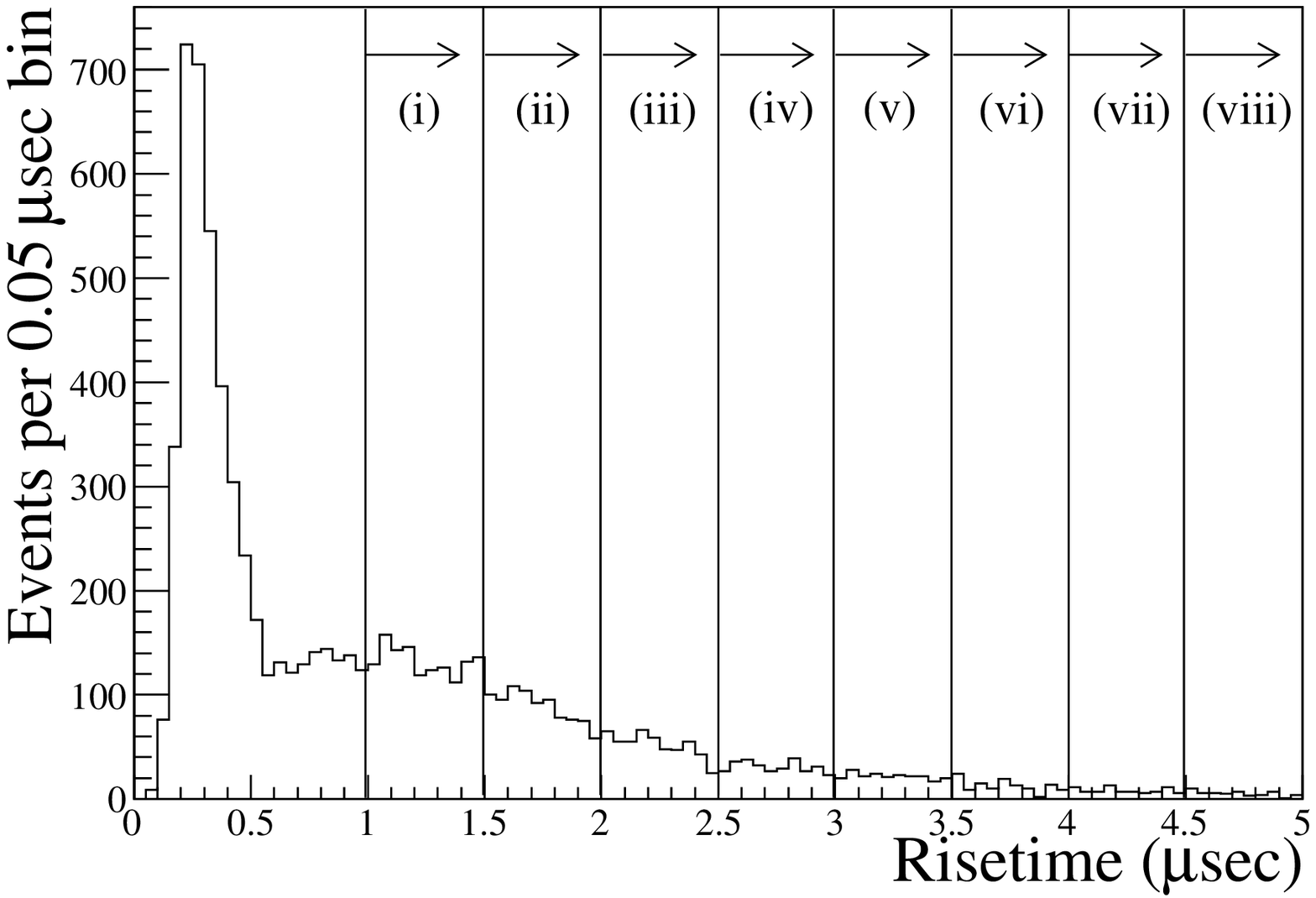,width=0.495\textwidth}}
       \subfigure[~Energy spectra for rise-time selection groups in (a).]{\epsfig{file=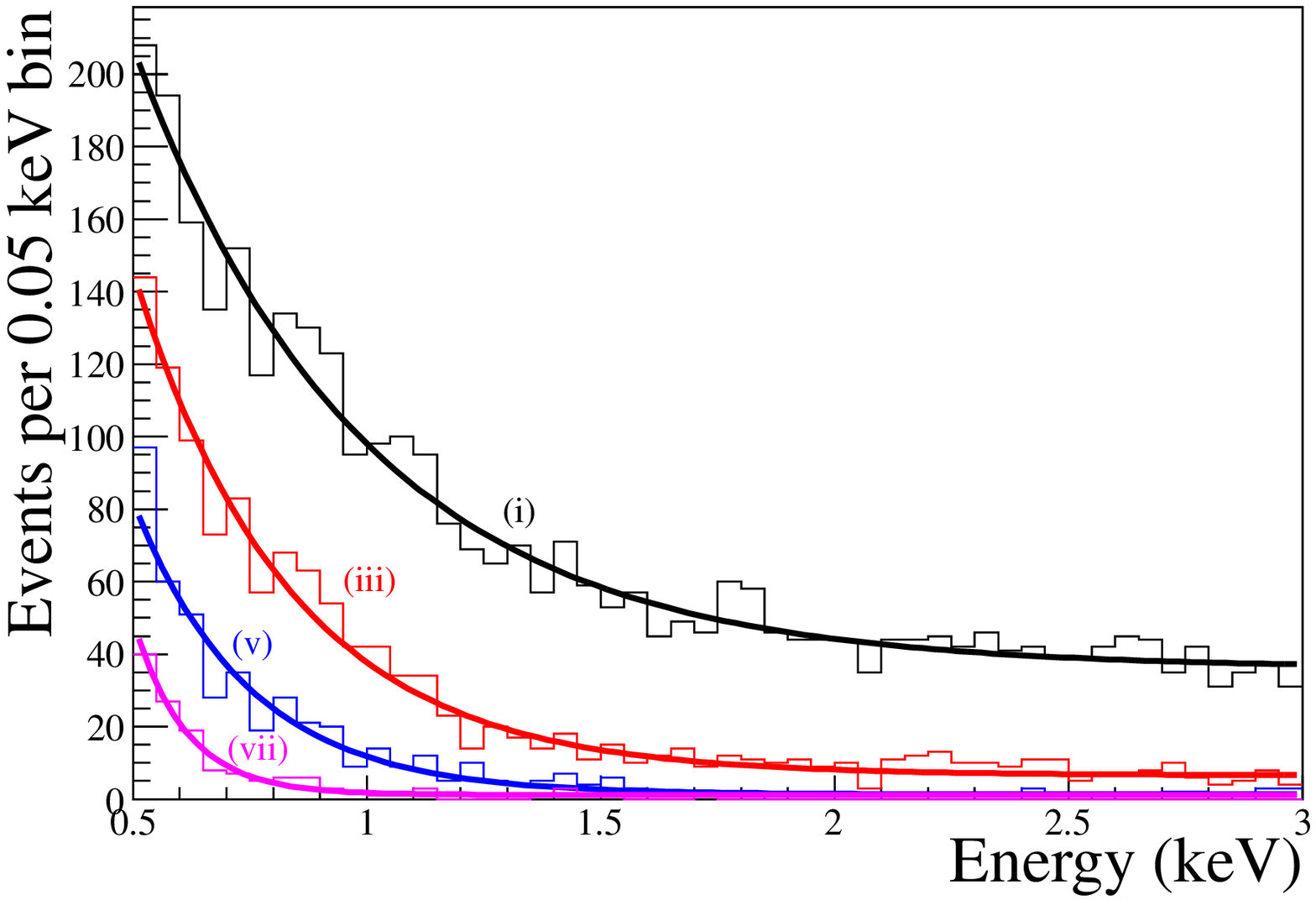,width=0.495\textwidth}}       
       \subfigure[~Exponential fit constants to energy spectra in (b).]{\epsfig{file=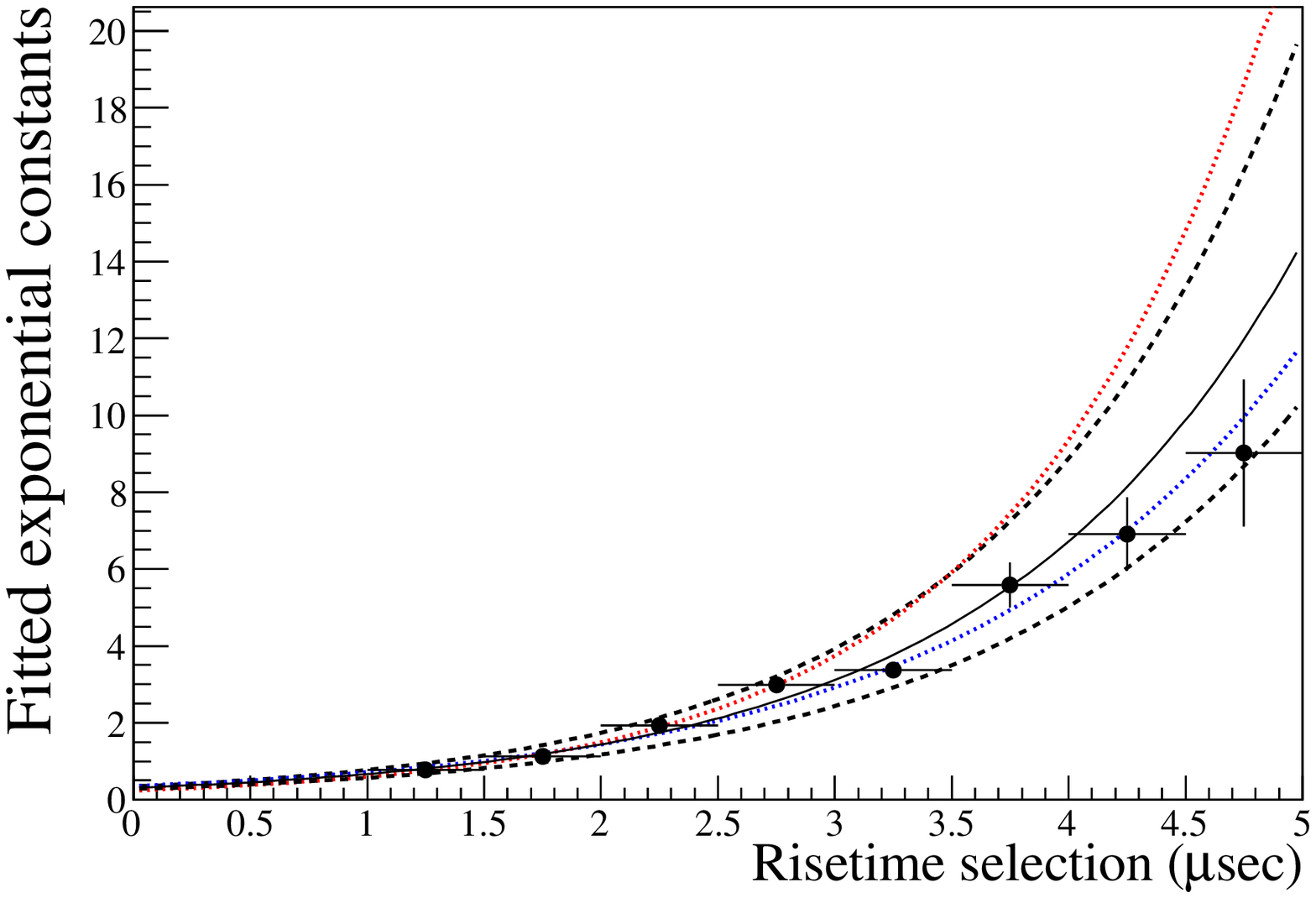,width=0.495\textwidth}}
       \subfigure[~Bounds on the energy distribution of surface events.]{\epsfig{file=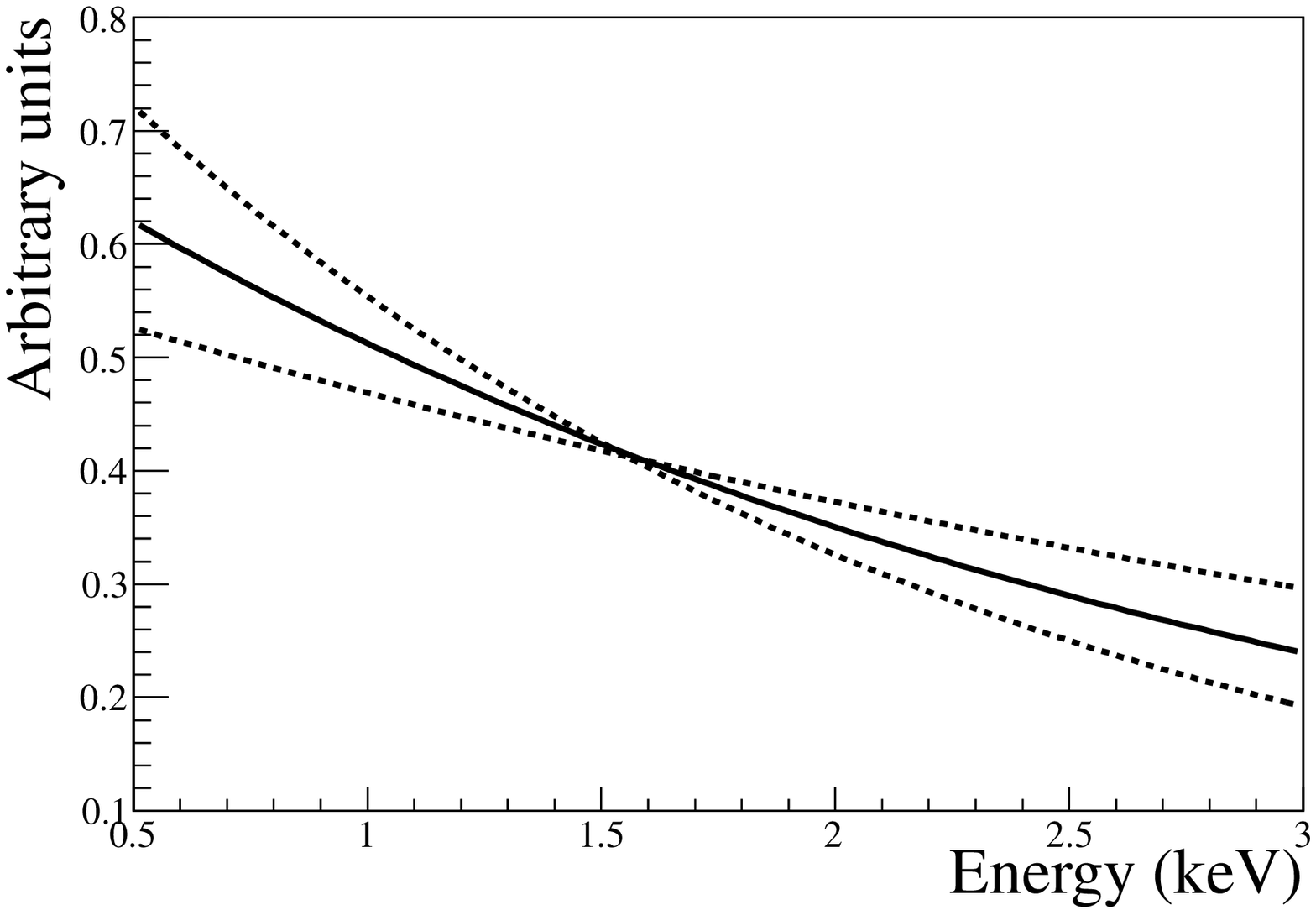,width=0.495\textwidth}}
       \caption[Steps taken to determine range of possible surface event energy distributions]{Steps taken to determine the systematic uncertainty on the surface event energy distribtution: (a) Select eight samples of surface events with different risetimes (risetime $>1$~$\mu$sec,  risetime $>1.5$~$\mu$sec, risetime $>2$~$\mu$sec, risetime $>2.5$~$\mu$sec, and so on), (b) Fit the corresponding energy distribution for each risetime selection to an exponential function, (c) Fit a functional form to the resulting exponential constants determined from step (b) to extrapolate the resulting exponential constant at risetime $>0$~$\mu$sec,  (d) Propagate uncertainties from step (c) to determine bounds on the surface event energy distribution.}
       \label{systudy}
    \end{center}
\end{figure*}

\begin{figure}[htbp]
   \includegraphics[width=0.5\textwidth]{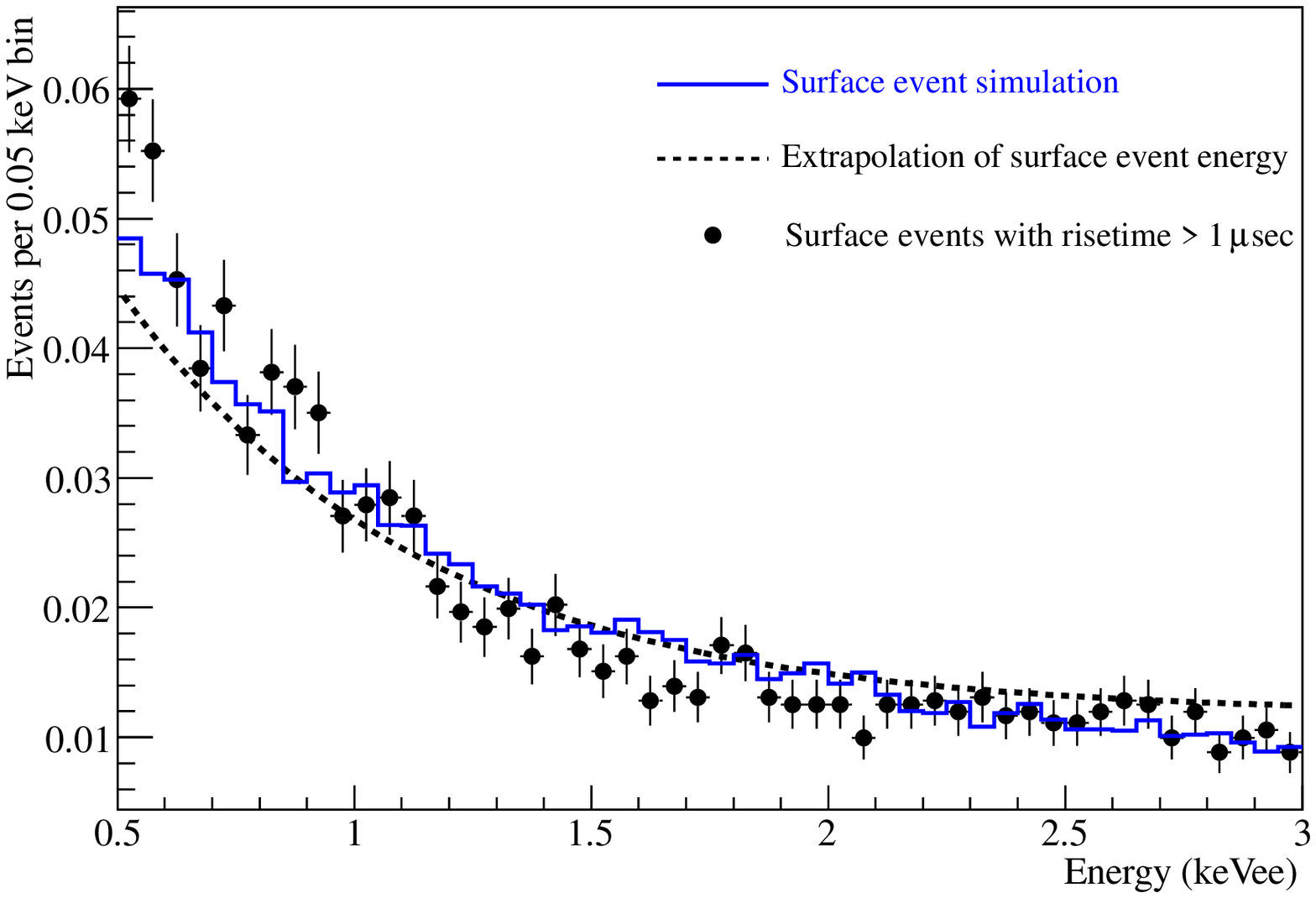}
   \caption{\label{mccompsurf} Comparision of the simulation of surface events from uranium and thorium chain gammas (blue solid line) to data with rise-time $>1$~$\mu$sec (black data points) and the extrapolation determined from Figure~\ref{systudy} (black dashed line).}
 \end{figure}

The largest uncertainty in the signal extraction comes from our understanding of the distributions of events that have energy deposition in the transition region near the surface of the crystal. This is the event population described in the maximum likelihood signal extraction formulation as $P_{\textrm{surf}}$.  Studies of surface events have been done on other detectors,  see for example \cite{ryan},~\cite{paddy}, and~\cite{CogPRD}.  However,  no surface event calibration has been done on the CoGeNT detector.  Thus, for our nominal surface event energy distribution we directly apply the transition region charge collection efficiency model described in~\cite{CogPRD} and~\cite{Sakai}.  To study the systematic uncertainty on this choice of energy distribution we look at high rise-time events in the CoGeNT detector to select a high-purity surface event sample.  The determination of the systematic uncerainty from the surface event energy distribution proceeds as follows:  (a) Select eight samples of surface events with different risetimes (risetime $>1$~$\mu$sec,  risetime $>1.5$~$\mu$sec, risetime $>2$~$\mu$sec, risetimeb$>2.5$~$\mu$sec, etc.), (b) Fit the corresponding energy distribution for each risetime selection to an exponential function,  (c) Fit a functional form to the resulting exponential constants determined from step (b) to extrapolate the resulting exponential constant at risetime $>0$~$\mu$sec, (d) Propagate uncertainties from step (c) to determine bounds on the surface event energy distribution.  We then perform the WIMP signal extraction using surface event distributions taken from both extrema of the energy distributions determined in the systemactic study.  Figure~\ref{systudy} illustrates the steps taken to determine the uncertainties on the surface event energy distribtution.   
As a validation of the surface event Monte Carlo, we compare the energy distribution of surface events taken from Monte Carlo to the extrapolated energy distribution taken from Figure~\ref{systudy}.  This comparison is shown in Figure~\ref{mccompsurf}.  While the comparison is not perfect, it shows the Monte Carlo and data are relativley close in terms of the surface event energy distributions.
   
We did not study the systematic uncertainty resulting from varying the thickness of the transition region in the Monte Carlo.  Varying the thickness of this region has a direct effect on the number of surface and bulk events.  However, the size of the surface event signal group is allowed to float in the signal extraction.

\section{Signal Extraction Results}
\label{sec:SigEx}
A likelihood signal extraction was performed to determine the magnitude of any possible WIMP dark matter signal in the low energy events (high-gain channel, 0.5--3~keVee) using the above described background PDFs.  A low-mass ($\sim$10~GeV/c$^{2}$) WIMP dark matter signal of high cross-section ($\sim$10$^{41}$~cm$^{2}$) would predominantly reside in this 0.5--3~keVee region. A standard halo model WIMP PDF signal extraction is first presented followed by several alternative PDF forms to explore the signal extraction method and the models of the CoGeNT data set.

\subsection{Signal extraction with the standard halo model}
\label{subsec:shm}
\begin{figure}[htbp]
   \includegraphics[width=0.5\textwidth]{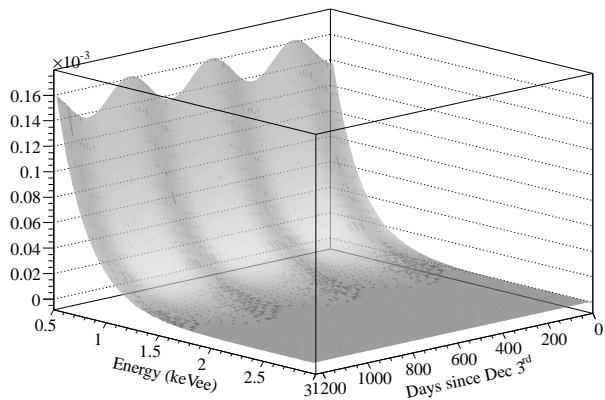}
   \caption{\label{fig:pdfs} The 2-D energy vs. time PDF used for the WIMP event category in the standard WIMP halo model signal extraction.}
\end{figure}

A signal extraction on the CoGeNT data was done using the standard WIMP halo model parameters of $v_{0} = 220$ km/s, $v_{esc} = $~550 km/s, and $\rho = $~0.3 GeV/$c^{2}$/cm$^{3}$.  In this extraction the 2-dimensional PDF shown in Figure~\ref{fig:pdfs} is used for the WIMP component in the extraction.
Of all the signal extractions performed in this analysis (see the following sections) the signal extraction with the standard halo model WIMP PDF had the least significance above the NULL result,  that is,  the case where no WIMP signal was included in the signal extraction.  In fact,  there was no difference in the Likelihood when including the WIMP signal compared to when it was not included.  Furthermore, there was no preference for any particular WIMP mass in the range tested (4 to 20 GeV).  The maximum number of WIMPs extracted in the standard WIMP halo model case was 5 events for the entire 1136 day dataset.  It can therefore be concluded that the data exhibits no preference for a standard halo model WIMP.

\subsection{Signal Extraction with free oscillation parameters}
\label{subsec:free}
\begin{figure}[!htbp]
   \begin{center}
   \subfigure[~Energy]{\epsfig{file=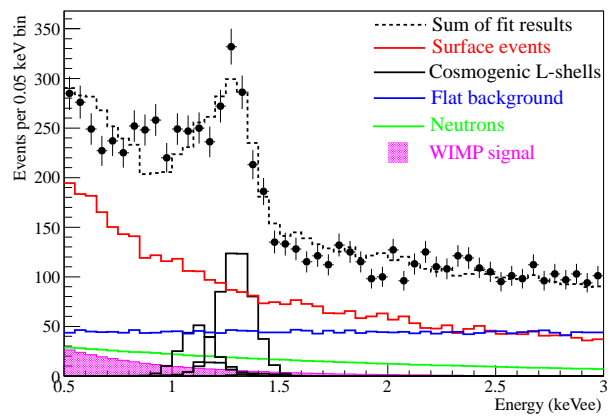,width=0.5\textwidth}}
   \subfigure[~Time]{\epsfig{file=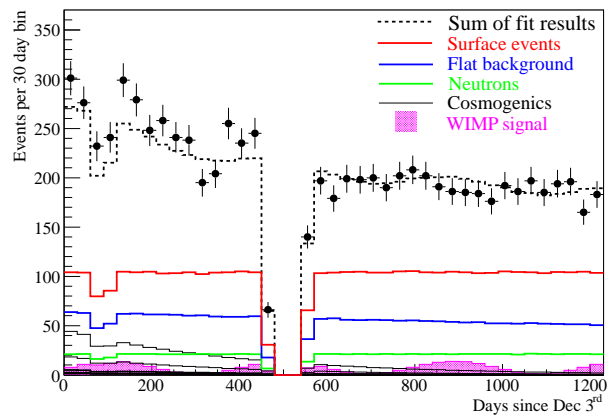,width=0.5\textwidth}}
   \caption{The extracted WIMP and background signals compared to the CoGeNT data in both energy and time.  The comparison in time is in 30 day bins.  This fit was performed with the WIMP oscillation amplitude, phase, and period all allowed to float.}
\label{timefit}
 \end{center}
 \end{figure}

To explore the possible existence of a non-standard halo model WIMP signal in the data a signal extraction was also performed where the WIMP oscillation parameters of amplitude, period, and phase were allowed to float.  The signal extraction with free oscillation parameters is strongly favored over the standard halo model.  However,  even in the most favorable signal extraction scenario for the existence of a possible WIMP signal in which the neutron PDF is constrained,  the WIMP signal is only favored over the NULL hypothesis at $<$ 2 $\sigma$.  Overall we found that the signal extractions that gave the best likelihood were when we let the gamma backgrounds be free to decay in time.  This includes both the bulk gamma and surface gamma backgrounds.  A decaying gamma background rate is used in the remainder of the signal extractions described. The best-fit half-life of the bulk Compton background (flat in energy) was 4143 $\pm$ 1812 days, and for the surface backgrounds it was 6424 $\pm$ 5140 days.  The half-life of the flat background is remarkably similar to the $^{3}$H half-life of 4500 days, though we also note that we do not believe $^{3}$H to be a large component of the flat background~\cite{CogPRD}.  The surface event background is similar to the $^{210}$Pb half-life of 8140 days, indicating that some fraction of the surface background may be due to $^{210}$Pb on the detector surface.  No conclusion is drawn from this observation due to the very large uncertainty on the extracted surface event half-life.  Figure~\ref{timefit} shows the fit results when we let the oscillation parameters float.  Panel (a) in figure~\ref{timefit} shows the fit results for this analysis compared to data in terms of energies while panel (b) shows the fit results with free WIMP oscillation parameters compared to the data in 30 day time bins.   The signal extraction results presented in the following subsections are summarized in table~\ref{sumtablewimp}.     

\subsubsection{Signal extraction with the neutron PDF constrained}
In the signal extraction where the magnitude of the neutron PDF is constrained we use the veto-germanium coincidence rate as a measure of the muon-induced neutron background.  This rate compares well with Monte carlo simulations of muon induced neutrons at the Soudan site~\cite{CogPRD}.  The way the constraint is applied in the signal extraction is shown in equation~\ref{econstraint}.  The resulting best-fit WIMP mass in this scenario is (13.0 $\pm$ 3.6) GeV/c$^{2}$.  The significance over the NULL result is only 1.7 $\sigma$.  An exclusion curve generated from this signal extraction is shown in figure~\ref{fig:CONT}.  The best fit oscillation parameters obtained from this extraction are a modulation amplitude of 0.67 $\pm$ 0.39,  period of 377 $\pm$ 20 days, and a phase of 97 $\pm$ 20 days.   
\subsubsection{Signal extraction with a free neutron PDF}
In another signal extraction the neutrons were allowed to float without any external constraints.  The best-fit WIMP mass in this case is (13.6 $\pm$ 2.6) GeV/c$^{2}$.  Again,  the signifance in this case is only 1.6 $\sigma$.  The exclusion curve generated from this signal extraction is also shown in figure~\ref{fig:CONT}.  The modulation parameters in this scenario are almost identical to those found in the neutron-constrained case.
\subsubsection{Signal extraction without a neutron PDF}
The signal extraction was also performed without neutrons included in the extraction.  The significance of this extraction over the NULL hypothesis is better,  however,  since there are no neutrons included in the extraction this does not represent a realistic background model.  We perform this extraction to provide the most conservative WIMP exclusion curve,  shown in figure~\ref{fig:CONT},  since in this case events that normally would be classified as neutrons are fitted as WIMPs.
\subsubsection{Signal extraction with fixed period and phase}
The final signal extraction performed was with the WIMP oscillation period fixed to 365 days and the phase fixed to 150 days.  The amplitude was still allowed to float.  In this case the significance of the fit was only 0.7 $\sigma$ above the NULL hypothesis.  This indicates that the period and phase \emph{expected} from WIMPs in the galactic halo are not supported by features observed in the data.

\begin{table*}[htbp]
\caption{\label{sumtablewimp} Summary of extracted WIMP mass and cross-section for the various signal extractions attempted.  Indicated in the table are the type of WIMP signal PDF used,  whether or not the neutron component of the signal extraction is constrained.}
\begin{ruledtabular}
\begin{tabular}{cccc}
\textrm{WIMP dark matter}&
\textrm{Neutron PDF}&
\textrm{Mass (GeV/c$^{2}$)}&
Null hypothesis\\
\textrm{PDF type}&
\textrm{constrained?}&
\textrm{ }&
\textrm{exclusion level}\\ \hline
Free oscillation parameters & Constrained & 13.0 & 1.7 $\sigma$ \\
Free oscillation parameters & Un-constrained & 13.6 & 1.6 $\sigma$ \\
Free oscillation parameters & fixed to 0 & 13.8 & 3.2 $\sigma$ \\
Fixed period and phase  & Constrained & 20 & 0.8 $\sigma$ \\
Fixed period and phase & Un-constrained & 14.8 & 0.7 $\sigma$ \\ 
\end{tabular}
\end{ruledtabular}
\end{table*}



\subsection{WIMP sensitivity}
\begin{figure}[!htbp]
   \includegraphics[width=0.5\textwidth]{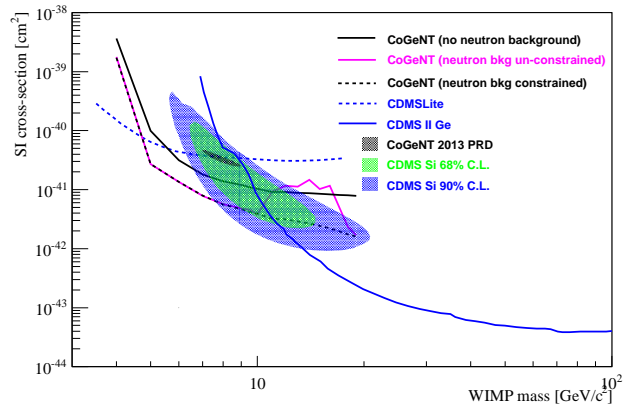}
   \caption{\label{fig:CONT}  Shown are the exclusion limits for various assumptions about the neutron component (see text).  Shown also are the 68\% C.L. and 90\% contours from the CDMS silicon result \cite{CDMSSi}.  The most conservative exclusion,  the one where we fix the number of neutrons to 0 in the fit,  can be taken as the most appropriate result of this analysis since no assumptions are made on the nature of the neutron background.}
 \end{figure}

Figure~\ref{fig:CONT} shows the WIMP sensitivity curves (2~$\sigma$ upper limits) derived from the likelihood analysis.   Only exclusion curves are drawn as all realistic background models fail to reject the NULL hypothesis at more than 3 $\sigma$.  The most conservative exclusion limit is determined from fixing the neutron component in the likelihood extraction to 0.  In this case,  there are more extracted WIMP events to compensate for the events that would be normally classified as neutrons,  therefore resulting in worse sensitivity.  

\section{Conclusions}
We have performed a likelihood signal extraction on the CoGeNT data using multi-dimensional PDFs for the backgrounds and a WIMP distribution.  The background PDFs are based on extensive Monte Carlo simulations done in \cite{CogPRD}.  In the signal extraction we use both the standard halo model and free WIMP oscillation parameters.  For all the signal extractions performed with realistic background models the NULL (no-WIMP) result is only excluded at $<$ 2 $\sigma$.  We also observe that depending on the PDF assumptions,  the results can have large variations.  We believe however that the method of a maximum likelihood signal extraction with floating background and signal PDFs is a powerful technique at extracting a WIMP signal.  Using this method we improve upon the previous CoGeNT WIMP sensitivity since we are better able to separate backgrounds from a possible WIMP signal.  This method becomes even more robust when the background distributions are well understood.  We conclude that this approach can be readily applied to other dark matter experiments especially in the low-mass WIMP region where a zero-background is often not achievable.     

\begin{acknowledgments}
We are indebted to Jeffrey de Jong and Alec Habig (MINOS collaboration) for sharing with us information on radon and muon rates at SUL, and to all SUL personnel for their constant support in operating the CoGeNT detector.  Work sponsored by the Pacific Northwest National Laboratory (PNNL) Ultra-Sensitive Nuclear Measurement Initiative LDRD program (Information Release number PNNL-SA-100496).  T.W.H. is partially supported by the Intelligence Community (IC) Postdoctoral Research Fellowship Program.
\end{acknowledgments}

\end{document}